\documentstyle[12pt]{article}
\newcommand{\beq}{\begin{equation}}
\newcommand{\eeq}{\end{equation}}
\newcommand{\bea}{\begin{eqnarray}}
\newcommand{\eea}{\end{eqnarray}}

\begin{document}

\title{{\bf N=2--Maxwell-Chern-Simons model with anomalous magnetic moment 
coupling via dimensional reduction }}
\author{{\normalsize {\bf H.R. Christiansen, M.S. Cunha, 
J.A. Helay\"{e}l-Neto}} \\
{\normalsize {\bf L.R.U. Manssur, A.L.M.A. Nogueira}}\\
\\
{\normalsize {\it Centro Brasileiro de Pesquisas F\'{\i}sicas, CBPF - DCP
\thanks{%
E-mail addresses: hugo, marcony, helayel, leon, nogue@cat.cbpf.br}}}\\
{\normalsize {\it Rua Dr. Xavier Sigaud 150, 22290-180 Rio de Janeiro, 
Brazil}}}
\maketitle

\begin{abstract}
{\normalsize \noindent
An N=1--supersymmetric version of the Cremmer-Scherk-Kalb-Ramond mo\-del
with non-minimal coupling to matter is built up both in terms of superfields
and in a component-field formalism. By adopting a dimensional reduction
procedure, the N=2--D=3 counterpart of the model comes out, with two main
features: a genuine (diagonal) Chern-Simons term and an anomalous magnetic
moment coupling between matter and the gauge potential.}
\end{abstract}

\newpage

\section{Introduction}

Ordinary and supersymmetric planar gauge models have been fairly well
investigated over the past years, in view of several remarkable properties
they exhibit. Among their most relevant features, we could quote:
gauge-invariant mass \cite{DJT}, ultraviolet finiteness \cite{piguet} and
the connection between extended supersymmetry and the existence of self-dual
soliton solutions \cite{marcony10/11}.

A few years ago, a Maxwell-Chern-Simons gauge theory with an additional
magnetic moment interaction was proposed \cite{marcony44}, for which
Bogomol'nyi-type self-dual equations can be derived and vortex-like
configurations appear whenever particular relations between the parameters
are obeyed \cite{navratil12/13/14}. An important issue that comes about is
the claim of a relation between the appearance of self-duality and the
N=2-supersymmetric extension of the model.

In this regard, Navr\'{a}til \cite{navratil} has succeeded in writing down
an N=2 Chern-Simons model with magnetic moment interaction. His paper relies
on a special choice of parameters in order that the supersymmetry be
extended. In our work, we also aim at an N=2 version of the
Maxwell-Chern-Simons model with magnetic moment interaction. However,
instead of building up our action directly in (1+2) dimensions and
constraining the parameters so as to achieve an N=2 extension as in \cite
{navratil}, we take the viewpoint of first formulating an N=1--D=4 gauge
model with a BF-term with no such constraints \cite{oswaldo}. Having in mind
a magnetic moment interaction in D=3, we consider matter non-minimally
coupled to a 2-form gauge potential in D=4 with completely independent
coupling constants (we refer to the latter as the Cremmer-Scherk-Kalb-Ramond
field). Upon a convenient dimensional reduction of the component-field
action from (1+3) to (1+2) dimensions, we set out an N=2--D=3 gauge model
with a Chern-Simons term and magnetic moment interaction with the matter
sector.

As we shall discuss later, our dimensional reduction procedure must be $\sup
$plemented by suitable field identifications that do not break the
supersymmetries of the extended model. This is necessary in order to ensure
that a genuine (non-mixed) Chern-Simons term drops out in 3 dimensions.

Our paper is outlined as follows. In Section 2, we propose the superfield
formulation of the N=1--D=4 gauge model with a BF--term and non-minimal
coupling between matter and the 2-form potential. Next, in Section 3, we
present the details of the dimensional reduction scheme we adopt. The
suitable field identifications, the N=2 transformations and the N=2--D=3
Maxwell-Chern-Simons action with anomalous magnetic moment interactions are
the subject of Section 4. Finally, in Section 5, we draw our General
Conclusions.


\section{The N=1--D=4 supersymmetric action}


We start off from the following superfield action:
\begin{equation}
{\cal S}_{4D}=\int d^4xd^2\theta \;\left\{ -\frac 18{\cal W}^a{\cal W}_a+d^2%
\overline{\theta }\;\left[ -\frac 12{\cal G}^2+\frac 12m{\cal VG}+\frac 1{16}%
\overline{\Phi }e^{2h{\cal V}}\Phi e^{4g{\cal G}}\right] \right\} ,
\label{eq1}
\end{equation}
where $m$ is a mass parameter, $h$ and $g$ are coupling constants, whereas ${%
\Phi ,}{\cal W}$ and ${\cal V}$ are superfields defined by the $\theta $%
-expansions below:
\begin{eqnarray}
\Phi &=&e^{(-i\theta \sigma ^\mu \overline{\theta }\partial _\mu )}\left(
\varphi (x)+\theta ^a\chi _a(x)+\theta ^2S(x)\right) ,\;\overline{D}_{\dot{a}%
}\Phi =0, \\
{\cal W}^a &=&-\frac 14\overline{D}^2D^a{\cal V},\;\;  \nonumber \\
{\cal V} &=&C(x)+\theta ^ab_a(x)+\overline{\theta }_{\dot{a}}\overline{b}^{%
\dot{a}}+\theta ^2H(x)+\overline{\theta }^2H^{*}(x)+\theta \sigma ^\mu
\overline{\theta }A_\mu +  \nonumber \\
&&\theta ^2\overline{\theta }\left( \overline{\lambda }-\frac i2\overline{%
\sigma }^\mu \partial _\mu b(x)\right) +\overline{\theta }^2\theta \left(
\lambda -\frac i2\sigma ^\mu \partial _\mu \overline{b}(x)\right) +\theta ^2%
\overline{\theta }^2\left( \triangle (x)-\frac 14\Box C(x)\right) .
\label{eq2}
\end{eqnarray}
Here $D_a$ and $\overline{D}_{\dot{a}}$ are the supersymmetric covariant
derivatives \cite{piguet/sibold}
\begin{eqnarray}
D_a &=&\partial _a-i\sigma _{\;a\dot{a}}^\mu \overline{\theta }^{\dot{a}%
}\partial _\mu  \nonumber \\
\overline{D}_{\dot{a}} &=&-\partial _{\dot{a}}+i\theta ^a\sigma _{\;a\dot{a}%
}^\mu \partial _\mu .
\end{eqnarray}
and ${\cal G}$ is defined in terms of the chiral spinor superfield
\begin{eqnarray}
\Sigma _a &=&\psi _a(x)+\theta ^b\Omega _{ba}(x)+\theta ^2\left[ {\xi
_a(x)+i\sigma _{a\dot{a}}^\mu \partial _\mu \overline{\psi }^{\dot{a}}(x)}%
\right] -i\theta \sigma ^\mu \overline{\theta }\partial _\mu \psi _a(x)
\nonumber \\
&&-i\theta \sigma ^\mu \overline{\theta }\theta ^b\partial _\mu \Omega
_{ba}(x)-\frac 14\theta ^2\overline{\theta }^2\Box \psi _a(x),\;\;\;\;\;%
\overline{D}_{\dot{a}}\Sigma _a=0,
\end{eqnarray}
by
\[
{\cal G}=\frac i8\left( D^a\Sigma _a-\overline{D}_{\dot{a}}\overline{\Sigma }%
^{\dot{a}}\right) .
\]

The Lorentz-group irreducible representations accommodated in $\Omega
_{ba}(x)$ can be split as follows:
\begin{equation}
\Omega _{ba}=\epsilon _{ba}\rho (x)+\left( \sigma ^{\mu \nu }\right) _{ba}%
{\cal B}_{\mu \nu }(x),
\end{equation}
with $\rho (x)$ and ${\cal B}_{\mu \nu }\left( x\right) $ being complex
fields:
\begin{eqnarray}
\rho (x) &=&P(x)+iM(x),\;  \nonumber \\
{\cal B}_{\mu \nu }(x) &=&\frac 14\left[ B_{\mu \nu }(x)-i\tilde{B}_{\mu \nu
}(x)\right] ,
\end{eqnarray}
with
\begin{equation}
\tilde{B}_{\mu \nu }(x)=\frac 12\varepsilon _{\mu \nu \alpha \beta
}B^{\alpha \beta }(x).
\end{equation}
In this way, ${\cal B}_{\mu \nu }$ exhibits a self-dual nature:
\begin{equation}
\widetilde{{\cal B}}_{\mu \nu }=i{\cal B}_{\mu \nu }.
\end{equation}
$B_{\mu \nu }$ is to be read as the 2-form field in the CSKR model which
emerges when one writes the action in components; therefore ${\cal G}$ is
referred to as the tensor multiplet\cite{alvaro}. The role of the remaining
field components introduced above will become clearer later. Let us mention
that the number of degrees of freedom is actually not as large as it seems.
For instance, both $\Phi $ and $\Sigma _a$ are chiral superfields, and the
superfield connection ${\cal V}$ will be taken in the Wess-Zumino(WZ) gauge
from now on,
\begin{equation}
{\cal V}=\theta \sigma ^\mu \overline{\theta }A_\mu +\theta ^2\overline{%
\theta }\overline{\lambda }+\overline{\theta }^2\theta \lambda +\theta ^2%
\overline{\theta }^2\triangle (x).
\end{equation}
However, for the sake of clarity, we have written it in its wider form, (\ref
{eq2}), because some of the susy variations shall explicitly exhibit the
compensating fields. As usual, an irreducible representation of the susy
algebra, involving just the field-strength $F_{\mu \nu }$ together with the
gaugino $\lambda $ and the auxiliary field $\Delta $ will be found. In fact,
the use of a complete expression for ${\cal V}$, as in eq.(\ref{eq2}), make
it easier to determine the transformation properties of the components under
susy transformations, and will enable us to find the proper identifications
necessary to formulate a Chern-Simons theory in 3D. Notice also that, within
the action, the spinor superfield comes into play only through the
field-strength superfield ${\cal G}$, which carries just half the degrees of
freedom of $\Sigma _a$: $\psi _a$ does not appear, $\rho $ appears only
through $M$, and on the same token ${\cal B}_{\mu \nu }$ manifests through $%
\tilde{G}_\mu $, making clear the resulting relevant degrees of freedom. The
component-field expansion for ${\cal G}$ turns out to be:
\begin{eqnarray}
{\cal G} &=&-\frac 12M+\frac i4\theta ^a\xi _a-\frac i4\bar{\theta}_{\dot{a}}%
\bar{\xi}^{\dot{a}}+\frac 12\theta ^a\sigma _{a\dot{a}}^\mu \bar{\theta}^{%
\dot{a}}\tilde{G}_\mu  \nonumber \\
&&+\frac 18\theta ^a\sigma _{a\dot{a}}^\mu \bar{\theta}^2\partial _\mu \bar{%
\xi}^{\dot{a}}-\frac 18\theta ^2\sigma _{a\dot{a}}^\mu \bar{\theta}^{\dot{a}%
}\partial _\mu \xi ^a-\frac 18\theta ^2\bar{\theta}^2\Box M,
\end{eqnarray}
where $G_{\mu \nu \kappa }$ and its dual, $\tilde{G}_\mu $, are given by
\begin{eqnarray}
G_{\alpha \mu \nu } &=&\partial _\alpha B_{\mu \nu }+\partial _\mu B_{\nu
\alpha }+\partial _\nu B_{\alpha \mu ,}  \nonumber \\
\tilde{G}_\mu &=&\frac 1{3!}\varepsilon _{\mu \nu \alpha \beta }G^{\nu
\alpha \beta }.
\end{eqnarray}
Therefore, the parametrization described above for ${\cal G}$ exhibits a
sort of WZ gauge effect for the spinor superpotential $\Sigma _a$ in that
the individual degrees of freedom carried by the latter can be grouped into
suitable combinations that correspond to the physical fields.%

Before looking at the action of eq.(\ref{eq1}) in terms of component fields,
notice that the coupling between matter and gauge fields exhibits the usual
exponential of the Maxwell superpotential ${\cal V}$, along with the
exponential of the superfield ${\cal G}.$ The latter has more consequences
than the former, since it carries gauge-invariant component fields which
shall appear to all orders in the action and cannot be reabsorbed upon field
redefinitions, as it is the case of the $C$--field appearing in the
expansion of ${\cal V}.$

Going over to components, the action $S_{4D}$ takes the form below:
\begin{eqnarray}
{\cal S}_{4D} &=&\int d^4x\left\{ -\frac 14F_{\mu \nu }F^{\mu \nu }+\frac
1{3!}G_{\mu \alpha \beta }G^{\mu \alpha \beta }+m\varepsilon ^{\mu \nu
\alpha \beta }A_\mu \partial _\nu B_{\alpha \beta }\right.  \nonumber \\
&&+2\Delta ^2+\frac i2\overline{\Lambda }\Gamma ^\mu \partial _\mu \Lambda
+\partial _\mu M\partial ^\mu M+\frac i4\overline{\Xi }\Gamma ^\mu \partial
_\mu \Xi +im\overline{\Lambda }\Gamma _5\Xi -4mM\Delta  \nonumber \\
&&+e^{-2gM(x)}\left[ \nabla _\mu \varphi \nabla ^\mu \varphi ^{*}+\frac i4%
\overline{X}\Gamma ^\mu \nabla _{\mu 5}X-\frac{g^2}2\partial _\mu M\left(
\overline{X}\Gamma _L\Gamma ^\mu \Xi \varphi ^{*}+\overline{\Xi }\Gamma
_L\Gamma ^\mu X\varphi \right) \right.  \nonumber \\
&&+\frac g2\left( \overline{\Xi }\Gamma ^\mu \Gamma _RX\nabla _\mu \varphi +%
\overline{X}\Gamma _L\Gamma ^\mu \Xi \nabla _\mu \varphi ^{*}\right) -i\frac{%
g^2}4\varphi ^{*}\varphi \overline{\Xi }\Gamma ^\mu \partial _\mu \Xi -\frac{%
g^2}{4h}\overline{\Xi }\Gamma _5\Gamma ^\mu {\cal J}_\mu \Xi  \nonumber
\label{scomp} \\
&&+\varphi \varphi ^{*}\left( 2h\Delta +igh\overline{\Lambda }\Gamma _5\Xi
-g^2\partial _\mu M\partial ^\mu M\right) -h(\varphi \overline{\Lambda }%
\Gamma _RX+\varphi ^{*}\overline{\Lambda }\Gamma _LX)  \nonumber \\
&&\left. \left. +\left( S-\frac{ig}2\overline{X}\Gamma _L\Xi +\frac{g^2}4%
\overline{\Xi }\Gamma _L\Xi \varphi \right) \left( S^{*}+\frac{ig}2\overline{%
X}\Gamma _R\Xi +\frac{g^2}4\overline{\Xi }\Gamma _R\Xi \varphi ^{*}\right)
\right] \right\} ,
\end{eqnarray}
where we have organized the fermionic fields so as to form four-component
Majorana spinors as follows:
\[
\Xi \left( x\right) \equiv \left(
\begin{array}{c}
\xi _a(x) \\
\bar{\xi}^{\dot{a}}(x)
\end{array}
\right) ,\;\;X\equiv \left(
\begin{array}{c}
\chi _a \\
\overline{\chi }^{\dot{a}}
\end{array}
\right) ,\;\;\Lambda \equiv \left(
\begin{array}{c}
\lambda _a \\
\overline{\lambda }^{\dot{a}}
\end{array}
\right) ,
\]
and the current ${\cal J}_\mu $ is given by
\begin{equation}
{\cal J}_\mu =-\frac{ih}2\left( \varphi ^{*}\nabla _\mu \varphi -\varphi
\nabla _\mu \varphi ^{*}\right) ,
\end{equation}
with
\begin{equation}
\nabla _\mu \varphi =\left( \partial _\mu +ihA_\mu +ig\tilde{G}_\mu \right)
\varphi .
\end{equation}
Also, there appears a covariant derivative with $\Gamma _5$-couplings
\begin{equation}
\nabla _{\mu 5}X=\left( \partial _\mu -ihA_\mu \Gamma _5-ig\tilde{G}_\mu
\Gamma _5\right) X.
\end{equation}
It is noteworthy to pay attention to the presence of the bosonic CSKR
Lagrangian among the first 3 terms of eq.(\ref{scomp}). We shall see in
Section 4 how the corresponding mixing term can be manipulated so as to give
rise to the usual Chern-Simons term. We can also recognize, in the first
term in the square brackets, a kinetic piece which corresponds to the
non-minimal coupling of scalar matter to the CSKR gauge fields. These four
terms define an Abelian gauge invariant theory which we will carefully
analyse as a guide to connect both gauge groups.

The scalar component field $M$ corresponds to a physical mode. It has
canonical dimension 1 (in units of mass) and it yields non-polynomial
interactions as it can be seen from the action ${\cal S}_{4D}$. This fact
destroys the renormalizability of the model. However, we stress that the $M$%
--field remains, contrary to what happens to the $C$--field present in the $%
{\cal V}$--superfield.

It is worthwhile noting that, in order to achieve the four-component spinors
in the expression above, we have chosen the following representation for the
$\Gamma $-matrices in (1+3) dimensions:
\[
\Gamma ^\mu =\left(
\begin{array}{cc}
0 & \sigma _{a\dot{b}}^\mu \\
\bar{\sigma}^{\mu \dot{a}b} & 0
\end{array}
\right) .
\]
Of course, the action is independent of such a choice and in the next
section we shall adopt a Majorana-like representation in order to perform
dimensional reduction.


The susy transformations of the components fields are listed below:
\begin{eqnarray}
\delta \varphi &=&\varepsilon ^a\chi _a,  \nonumber \\
\delta \chi _a &=&2\varepsilon _aS-2i\sigma _{a\dot{a}}^\mu \bar{\varepsilon}%
^{\dot{a}}D_\mu \varphi , \\
\delta S &=&-i\bar{\varepsilon}_{\dot{a}}\bar{\sigma}^{\mu \dot{a}a}D_\mu
\chi _a+2h\overline{\lambda }_{\dot{a}}\bar{\varepsilon}^{\dot{a}}\varphi ;
\nonumber
\end{eqnarray}
\begin{eqnarray}
\delta M &=&\frac i2\bar{\varepsilon}_{\dot{a}}\bar{\xi}^{\dot{a}}-\frac
i2\varepsilon ^a\xi _a,  \nonumber \\
\delta \xi _a &=&2\sigma _{a\dot{a}}^\mu \overline{\varepsilon }^{\dot{a}%
}\left( \partial _\mu M-i\tilde{G}_\mu \right) , \\
\delta \tilde{G}^\mu &=&\frac i2\varepsilon ^b\left( \sigma ^{\mu \nu
}\right) _b^{\;\;a}\partial _\nu \xi _a+\frac i2\bar{\varepsilon}_{\dot{b}%
}\left( \bar{\sigma}^{\mu \nu }\right) _{\;\dot{a}}^{\dot{b}}\partial _\nu
\bar{\xi}^{\dot{a}};  \nonumber
\end{eqnarray}
\begin{eqnarray}
\delta A^\mu &=&\varepsilon ^a\sigma _{a\dot{a}}^\mu \overline{\lambda }^{%
\dot{a}}-\bar{\varepsilon}_{\dot{a}}\bar{\sigma}^{\mu \dot{a}a}\lambda _a,
\nonumber \\
\delta \lambda _a &=&2\varepsilon _a\Delta +\frac i2\sigma _{a\dot{b}}^\nu
\overline{\sigma }^{\mu \dot{b}b}\varepsilon _bF_{\mu \nu }, \\
\delta \Delta &=&-\frac i2\varepsilon ^a\sigma _{a\dot{a}}^\mu \partial _\mu
\bar{\lambda}_{\dot{a}}-\frac i2\bar{\varepsilon}_{\dot{a}}\bar{\sigma}^{\mu
\dot{a}a}\partial _\mu \lambda _a.  \nonumber
\end{eqnarray}
Now, it is clear that the first and the second groups form respectively two
irreducible representations of the susy algebra while the last three terms,
together with
\begin{equation}
\delta F_{\mu \nu }=\varepsilon ^a\sigma _{[\nu a\dot{a}}\partial _{\mu ]}%
\overline{\lambda }^{\dot{a}}-\bar{\varepsilon}_{\dot{a}}\bar{\sigma}_{[\nu
}^{\dot{a}a}\partial _{\mu ]}\lambda _a,
\end{equation}
close another one. On the other hand, $A_\mu $ transforms along with the
compensating fields $b$ and $\overline{b}$ but we can always fix the WZ
gauge to eliminate them. In fact, it is not worth exhibiting the susy
transformations of any of the compensating fields themselves as they become
zero in the WZ gauge. Analogously, in the second group of variations, it can
be seen that (as it occurs in the action) $P$, the real part of $\rho $, is
irrelevant, $\psi _a$ does not appear and $B_{\mu \nu }$ contributes only
through $\tilde{G}_\mu $. We have however shown the variation of $A_\mu $
just because it enters the action not exclusively through $F_{\mu \nu }$.


The list of field variations in terms of four-component spinors is then
\begin{eqnarray}
\delta \varphi &=&\overline{{\cal E}}\Gamma _LX  \nonumber \\
\delta X &=&2\left( S-i\Gamma ^\mu D_\mu \varphi ^{*}\right) \Gamma _L{\cal E%
}+2\left( S^{*}-i\Gamma ^\mu D_\mu \varphi \right) \Gamma _R{\cal E} \\
\delta S &=&-i\overline{{\cal E}}\Gamma ^\mu \Gamma _LD_\mu X+2h\overline{%
{\cal E}}\Gamma _R\Lambda \varphi  \nonumber
\end{eqnarray}
\begin{eqnarray}
\delta M &=&\frac i2\overline{{\cal E}}\Gamma _5\Xi  \nonumber \\
\delta \Xi &=&2\Gamma ^\mu \left( \Gamma _5\partial _\mu M-i\tilde{G}_\mu
\right) {\cal E} \\
\delta \tilde{G}^\mu &=&\frac i2\overline{{\cal E}}\Gamma ^{\mu \nu
}\partial _\nu \Xi  \nonumber
\end{eqnarray}
\begin{eqnarray}
\delta F_{\mu \nu } &=&\overline{{\cal E}}\Gamma ^{[\nu }\Gamma _5\partial
^{\mu ]}\Lambda  \nonumber \\
\delta \Lambda &=&2\Delta {\cal E}-\frac i2\Gamma _5\Gamma ^\mu \Gamma ^\nu
F_{\nu \mu }{\cal E} \\
\delta \Delta &=&-\frac i2\overline{{\cal E}}\Gamma ^\mu \partial _\mu
\Lambda ,  \nonumber
\end{eqnarray}
${\cal E}$ being the (infinitesimal) Majorana-spinor parameter of the susy
transformation,
\[
{\cal E}\equiv \left(
\begin{array}{c}
\varepsilon _a \\
\bar{\varepsilon}^{\dot{a}}
\end{array}
\right) .\;\;
\]


\section{The dimensional reduction: from D=4 to D=3}


From now on, we shall identify four-dimensional Lorentz indices by $\hat{\mu}%
=0,1,2,3,$ while in three-dimensional space-time we will keep bare greek
indices, namely, $\mu =0,1,2.$

Let us first perform the dimensional reduction of the bosonic sector of the
susy action, eq.(\ref{scomp}). For this, we will adopt the following
procedure: we eliminate the third spatial coordinate so as to make the
D=3-fields $x_3$-independent \cite{sherk},
\begin{equation}
\partial _3\mbox{(fields)}=0.  \label{x3}
\end{equation}
On the other hand, we will assume that the $\hat{\mu}=3$ component of the
D=4-fields are taken as scalars in (1+2) dimensions. In this scheme, the
Poincar\'{e} invariance in D=1+3 has been broken down to the direct product
between Poincar\'{e} invariance in D=1+2 and a U(1) factor. Thus, $A_\mu $
is in the vector representation of the Lorentz group and is a singlet of
such a U(1) while $A_3$ is an independent scalar field. The relevant
(off-shell) degrees of freedom of $B_{\hat{\mu}\hat{\nu}}$ are $B_{\mu \nu }$
and $B_{\mu 3}$ as it is an antisymmetric tensor. Accordingly, in
three-dimensional space, we shall make the following identifications:
\begin{eqnarray}
&&N\equiv A^3,\ \ B^\mu \equiv B^{3\mu },\ \ B^{\mu \nu }\equiv \varepsilon
^{\mu \nu \rho }Z_\rho ,  \nonumber \\
&&\partial _\mu Z^\mu =-\tilde{G}^3,\ \ \partial ^\mu B^\nu -\partial ^\nu
B^\mu =G^{\mu \nu },
\end{eqnarray}
with $\varepsilon ^{\mu \nu \rho }\equiv \varepsilon ^{\mu \nu \rho 3}$.
Hence, after decomposition, the bosonic terms reduce as given below:
\begin{eqnarray*}
-\frac 16G_{\hat{\mu}\hat{\nu}\hat{\rho}}G^{\hat{\mu}\hat{\nu}\hat{\rho}}
&\longrightarrow &-\frac 12G_{\mu \nu }G^{\mu \nu }+\partial _\mu Z^\mu
\partial _\nu Z^\nu , \\
-\frac 14F_{\hat{\mu}\hat{\nu}}F^{\hat{\mu}\hat{\nu}} &\longrightarrow
&-\frac 14F_{\mu \nu }F^{\mu \nu }+\frac 12\partial _\mu N\partial ^\mu N, \\
m\varepsilon _{\hat{\mu}\hat{\nu}\hat{\rho}\hat{\lambda}}A^{\hat{\mu}%
}\partial ^{\hat{\nu}}B^{\hat{\rho}\hat{\lambda}} &\longrightarrow
&2m\varepsilon _{\mu \nu \rho }A^\mu \partial ^\nu B^\rho +2mN\partial _\mu
Z^\mu , \\
\nabla _{\hat{\mu}}\phi ^{*}\nabla ^{\hat{\mu}}\phi &\longrightarrow &\nabla
_\mu \phi ^{*}\nabla ^\mu \phi -\left( hN-g\partial _\mu Z^\mu \right)
^2\left| \varphi \right| ^2.
\end{eqnarray*}

In order to proceed with the dimensional reduction in the fermionic sector,
let us mention that one can always construct a representation of the
Clifford algebra in the form of a tensor product of lower dimensional
matrices. We use capital $\Gamma ^{\hat{\mu}}$ for Dirac matrices in the
higher dimension and lower case $\gamma ^\mu $ in the lower dimension. A
suitable set of the 4D $\Gamma $-matrices is the following
\[
\Gamma ^\mu =\left(
\begin{array}{cc}
\gamma ^\mu & 0 \\
0 & \mbox{-}\gamma ^\mu
\end{array}
\right) ,\ \ \ \ \ \ \ \ \Gamma ^3=\left(
\begin{array}{cc}
0\ \  & i \\
i\ \  & 0
\end{array}
\right) .
\]
Taking $\gamma ^0\equiv \sigma _y,\;\gamma ^1\equiv i\sigma _x,\;$and $%
\gamma ^2\equiv i\sigma _z,$ we have a Majorana representation both in D=4
and D=3. So, in this way, a Majorana spinor in D=4 is real and splits into a
doublet of real Majorana spinors in D=3.

It is worth noting, before dimensionally reducing the fermionic sector, that
the relevant degrees of freedom reorganize themselves as follows. The field
content of four-component spinors:
\begin{eqnarray*}
X &\rightarrow &\chi ,\;\omega \\
\Xi &\rightarrow &\xi ,\;\zeta \\
\Lambda &\rightarrow &\lambda ,\;\eta
\end{eqnarray*}
gives rise to the following Dirac spinors
\begin{eqnarray}
X_{\pm } &=&\chi \pm i\omega  \nonumber \\
\Xi _{\pm } &=&\xi \pm i\zeta  \label{dirac3} \\
\Lambda _{\pm } &=&\lambda \pm i\eta  \nonumber
\end{eqnarray}
Namely, the two-component Majorana fermions corresponding to each 4D spinor
become completely independent in 3D and, further, in the reduced action they
appear as Dirac spinors in the particular way indicated in eq.(\ref{dirac3}%
). On the same footing, the infinitesimal susy parameter will break down
into two dissociated spinorial species
\begin{equation}
{\cal E}\longrightarrow \varepsilon ,\delta \ \rightarrow \varepsilon _{\pm
}=\varepsilon \pm i\delta ,
\end{equation}
revealing the existence of two supersymmetries in the reduced theory.


\section{\strut The N=2--D=3 Model}


In terms of the D=3 bosonic fields and Dirac fermions defined above, the
three-dimension\-al action reads
\begin{eqnarray}
{\cal S}_{3D} &=&\int d^3x\left\{ -\frac 14F_{\mu \nu }F^{\mu \nu }-\frac
12G_{\mu \nu }G^{\mu \nu }+2m\varepsilon ^{\mu \nu \alpha }A_\mu \partial
_\nu B_\alpha +2\Delta ^2+(\partial _\mu Z^\mu )^2\right.  \nonumber \\
&&+\frac i2\overline{\Lambda }_{-}\partial \!\!\!/\,\Lambda _{-}+\frac i4%
\overline{\Xi }_{-}\partial \!\!\!/\,\Xi _{-}+\frac i2m(\overline{\Lambda }%
_{+}\Xi _{-}-\overline{\Lambda }_{-}{\Xi }_{+})+\frac 12\partial _\mu
N\partial ^\mu N  \nonumber \\
&&-4mM\Delta +2mN\partial _\mu Z^\mu +\partial _\mu M\partial ^\mu
M+e^{-2gM}\left[ \nabla _\mu \varphi \nabla ^\mu \varphi ^{*}-(hN-g\partial
_\mu Z^\mu )^2\,\varphi \varphi ^{*}\right.  \nonumber \\
&&+\frac 14(hN-g\partial _\mu Z^\mu )\,\overline{X}_{+}X_{+}+\frac i8(%
\overline{X}_{-}\nabla \!\!\!\!/\,_{-}X_{-}+\overline{X}_{+}\nabla
\!\!\!\!/\,X_{+})  \nonumber \\
&&+\frac g4\left[ \left( \overline{\Xi }_{-}\gamma ^\mu X_{-}+\overline{X}%
_{+}\gamma ^\mu \Xi _{+}\right) \nabla _\mu \varphi -i\left( \overline{\Xi }%
_{-}X_{-}\varphi +\overline{X}_{-}\Xi _{-}\varphi ^{*}\right) (hN-g\partial
_\mu Z^\mu )\right]  \nonumber \\
&&-\frac{g^2}4\partial _\mu M\left( \overline{X}_{-}\gamma ^\mu \Xi
_{-}\varphi ^{*}+\overline{\Xi }_{-}\gamma ^\mu X_{-}\varphi \right) -i\frac{%
g^2}8\varphi ^{*}\varphi \left( \overline{\Xi }_{-}\partial \!\!\!/\,\Xi
_{-}+\overline{\Xi }_{+}\partial \!\!\!/\,\Xi _{+}\right)  \nonumber \\
&&+\frac{g^2}{4h}\left( \frac 12(\overline{\Xi }_{-}\gamma ^\mu {\cal J}_\mu
\Xi _{-}-\overline{\Xi }_{+}\gamma ^\mu {\cal J}_\mu \Xi _{+})+\overline{\Xi
}_{-}\Xi _{-}h(hN-g\partial _\mu Z^\mu )\varphi \varphi ^{*}\right)
\nonumber  \label{scomp3} \\
&&+\varphi \varphi ^{*}\left( 2h\Delta +\frac{igh}2(\overline{\Lambda }%
_{+}\Xi _{-}-\overline{\Xi }_{-}\Lambda _{+})-g^2\partial _\mu M\partial
^\mu M\right) -\frac h2(\varphi \overline{\Lambda }_{+}X_{-}+\varphi ^{*}%
\overline{X}_{-}\Lambda _{+})  \nonumber \\
&&\left. \left. +\left| S-\frac{ig}4\overline{X}_{-}\Xi _{+}+\frac{g^2}%
8\varphi \overline{\Xi }_{-}\Xi _{+}\right| ^2\right] \right\} .
\label{S-3d}
\end{eqnarray}
Although it looks so large an expression, it has been written in a rather
compact notation and, again, it can be recognized that it contains a mixed
CS theory. Of course, we would like to know more about its physical meaning,
which is still fairly obscure. Actually, the bosonic sector characterizes a
parity-preserving statistical gauge theory which can be related to
superconductivity at finite temperature \cite{Dorey(QED3 4)}. At this point,
we should draw the attention to the $Z_\mu $--field, dual of the two-form $B$
in three dimensions. Its kinetic term is not built up from the usual field
strength, as it is the case for ordinary gauge vector fields. An inspection
of its Abelian transformation shows that its transverse part can be gauged
away. This is why only its longitudinal part propagates off-shell. Such a
peculiar gauge field does not correspond to any physical excitation: a
two-form gauge field presents no on-shell degree of freedom in $D=3$. So,
the kinetic term for Z$_\mu $ is harmless, for no ghost excitation is
present in the spectrum.

In the next section, we shall find a remarkable result coming from a
detailed inspection of the Lagrangian (\ref{S-3d}) and the susy
transformations, suggesting a simple identification between some of the
several fields appearing at the present stage. Indeed, as we shall propose
later, the 3-divergence of $Z_\mu $ will be identified as the auxiliary
component of the gauge superfield. Now, let us evaluate the two susy
transformations acting on the 3D fields. The scalar multiplet transforms as
\begin{eqnarray}
\delta \varphi &=&\frac 12\bar{\varepsilon}_{-}X_{+},  \nonumber \\
\delta S &=&-\frac i2\bar{\varepsilon}_{+}\left( D\!\!\!\!/\,-ihN\right)
X_{+}+h\bar{\varepsilon}_{+}\Lambda _{-}\varphi , \\
\delta X_{+} &=&2S\varepsilon _{+}+2\left( hN\varphi -iD\!\!\!\!/\,\varphi
\right) \varepsilon _{-},  \nonumber \\
\ \delta X_{-} &=&2S^{*}\varepsilon _{-}+2\left( hN\varphi
^{*}-iD\!\!\!\!/\,\varphi ^{*}\right) \varepsilon _{+};  \nonumber
\end{eqnarray}
the vector multiplet transformations read as follows 
\begin{eqnarray}
\delta N &=&-\frac 12(\bar{\varepsilon}_{+}\Lambda _{+}+\bar{\varepsilon}%
_{-}\Lambda _{-}),  \nonumber \\
\delta \Lambda _{\pm } &=&(2\Delta +i\partial \!\!\!/\,N)\varepsilon _{\pm
}\pm \gamma _\mu \tilde{F}^\mu \varepsilon _{\pm },  \nonumber \\
\delta \Delta &=&-\frac i2(\bar{\varepsilon}_{+}\partial \!\!\!/\,\Lambda
_{+}+\bar{\varepsilon}_{-}\partial \!\!\!/\,\Lambda _{-}), \\
\delta \tilde{F}^\mu &=&-\frac 12(\bar{\varepsilon}_{+}\varepsilon ^{\mu \nu
\lambda }\gamma _\lambda \partial _\nu \Lambda _{+}-\bar{\varepsilon}%
_{-}\varepsilon ^{\mu \nu \lambda }\gamma _\lambda \partial _\nu \Lambda
_{-}),  \nonumber
\end{eqnarray}
and, finally, the tensor multiplet components transform according to 
\begin{eqnarray}
\delta M &=&\frac i4(\bar{\varepsilon}_{+}\Xi _{-}-\bar{\varepsilon}_{-}\Xi
_{+}),  \nonumber \\
\delta \Xi _{\pm } &=&\pm 2(\partial \!\!\!/\,M+i\partial _\mu Z^\mu
)\varepsilon _{\mp }-2i\gamma _\mu \tilde{G}^\mu \varepsilon _{\mp },
\nonumber \\
\delta \tilde{(}\partial _\mu Z^\mu ) &=&\frac 14(\bar{\varepsilon}%
_{+}\,\partial \!\!\!/\Xi _{-}-\bar{\varepsilon}_{-}\partial \!\!\!/\,\Xi
_{+}), \\
\delta \tilde{G}^\mu &=&-\frac i4(\bar{\varepsilon}_{-}\varepsilon ^{\mu \nu
\lambda }\gamma _\lambda \partial _\nu \Xi _{+}+\bar{\varepsilon}%
_{+}\varepsilon ^{\mu \nu \lambda }\gamma _\lambda \partial _\nu \Xi _{-}).
\nonumber
\end{eqnarray}
These susy variations show on the one hand that $S_{3D}$ is indeed
N=2--supersymmetric, and, on the other hand, they exhibit the key to get
one's hands on the underlying Chern-Simons Lagrangian by making manifest the
relevant degrees of freedom to realize the supersymmetry algebra.


Once the N=2 transformation laws have been cleared up, we shall demonstrate
that the previous susy action (\ref{scomp3}) may be suitably manipulated so
as to give rise to a more familiar system, namely, the supersymmetric
extension of a non-minimal Maxwell-Chern-Simons theory.

The first natural attempt is to associate the two vector gauge fields and
then look for the corresponding fermionic connection in order to keep both
supersymmetries. Doing so, one gets to the conclusion that, by means of
simply identifying $A_\mu $ and $B_\mu \;\left( A_\mu \equiv B_\mu \right) $%
, we can complete the remaining identifications:
\begin{equation}
N\equiv -M,\ \ \ \Delta \equiv -\frac 12\tilde{G}^3,\ \ \ \Lambda _{\pm
}\equiv \pm \frac i2\Xi _{\mp }  \label{ident}
\end{equation}
so as to achieve the connection between the two gauge groups of the outset
and obtain a proper MCS N=2-supersymmetric theory with non-minimal coupling.
As long as the two sets of fields transform identically under a symmetry
transformation, we may identify them without breaking supersymmetry.

The identification between $A_\mu $ and $B_\mu $ is to be regarded as part
of our ansatz for dimensional reduction. We take this viewpoint having in
mind to achieve a genuine diagonal MCS term. This can be performed at the
level of the action provided that no inconsistency shows up at the level of
the supersymmetry transformations for the component fields. This has
actually been checked together with the identifications displayed in eq.(\ref
{ident}): the supersymmetry algebra turns out to be consistent with this
ansatz. We should perhaps point out that the non-identification of $A_\mu $
and $B_\mu $ yields a mixed MCS term and the spectrum displays two physical
modes: a massless and a massive vector. So, the reduction, with and without
the identification of these potentials, leads to two non-equivalent models.
Our ansatz works as a mapping between them.

There is still another point deserving a comment in connection with this
field identification: the original 4-dimensional model exhibits two $U(1)$%
--factors. The emergence of the extra vector potential $B_\mu $ in the
reduced model triggers an extra $U(1)$--symmetry, whose scalar parameter
comes out from the reduction of the vector parameter associated to the
Abelian symmetry of the 2--form potential in four dimensions. Had we not
identified $A_\mu $ and $B_\mu $ the 3--dimensional model would present a $%
[U(1)]^3$--symmetry. With our ansatz, our symmetry for the reduced model is
still a $U(1)$x$U(1)$--invariance. Clearly, the field identifications
imposes contemporarily a corresponding identification of the gauge
parameters associated to the Abelian symmetries attached to $A_\mu $ and $%
B_\mu .$ As a matter of fact, the residual symmetry of the reduced model is
indeed a single $U(1)$--symmetry, for the gauge potential $Z_\mu $ can be
completely gauged away and the matter fields do not have charge relative to
the $U\left( 1\right) $--invariance associated to this gauge potential.

In order to get the D=3-action with the canonical kinetic terms it is
convenient to redefine the superfields and coupling constants of eq.(\ref
{eq1}) as follows
\begin{eqnarray*}
&&{\cal V}\rightarrow {\cal V}/\sqrt{3}\ \ \ {\cal G}\rightarrow {\cal G}/%
\sqrt{3} \\
&&h\rightarrow \sqrt{3}h\ \ \ g\rightarrow \sqrt{3}g\ \ \ m\rightarrow \frac
34m.
\end{eqnarray*}
In doing so, the action of eq.(\ref{eq1}) becomes
\begin{equation}
{\cal S}_{4D}=\int d^4xd^2\theta \;\left\{ -\frac 1{24}{\cal W}^a{\cal W}%
_a+d^2\overline{\theta }\;\left[ -\frac 16{\cal G}^2+\frac 18m{\cal VG}%
+\frac 1{16}\overline{\Phi }e^{2h{\cal V}}e^{4g{\cal G}}\Phi \right] \right\}
\label{eq1new}
\end{equation}
bringing about the usual MCS terms in the N=2-susy action: 
\begin{eqnarray}
{\cal S}_{MCS}^{N=2} &=&\int d^3x\left\{ -\frac 14F_{\mu \nu }F^{\mu \nu
}+\frac m2\varepsilon ^{\mu \nu \alpha }A_\mu \partial _\nu A_\alpha
+2\Delta ^2+\frac 12\partial _\mu N\partial ^\mu N+2mN\Delta \right.
\nonumber \\
&&+\frac 12\overline{\Lambda }_{-}(i\partial \!\!\!/\,+m)\Lambda
_{-}+e^{2gN}\left[ \nabla _\mu \varphi \nabla ^\mu \varphi ^{*}-(hN-2g\Delta
)^2\,\varphi \varphi ^{*}\right.  \nonumber \\
&&+\frac 14(hN-2g\Delta )\,X_{+}X_{+}+\frac i8(\overline{X}_{-}\nabla
\!\!\!\!/\,_{-}X_{-}+\overline{X}_{+}\nabla \!\!\!\!/\,X_{+})  \nonumber \\
&&+\frac{ig}2\left( \overline{\Lambda }_{+}\nabla \!\!\!\!/\,\varphi
X_{-}-h.c.\right) -\frac g2(hN-2g\Delta )\left( \overline{\Lambda }%
_{+}X_{-}\varphi +h.c.\right)  \nonumber \\
&&-\frac{ig^2}2\partial _\mu N\left( \overline{X}_{-}\gamma ^\mu \Lambda
_{+}\varphi ^{*}-\overline{\Lambda }_{+}\gamma ^\mu X_{-}\varphi \right) -i%
\frac{g^2}2\varphi ^{*}\varphi \left( \overline{\Lambda }_{+}\partial
\!\!\!/\,\Lambda _{+}+\overline{\Lambda }_{-}\partial \!\!\!/\,\Lambda
_{-}\right)  \nonumber \\
&&+\frac{g^2}h\left( \frac 12(\overline{\Lambda }_{+}\gamma ^\mu {\cal J}%
_\mu \Lambda _{+}-\overline{\Lambda }_{-}\gamma ^\mu {\cal J}_\mu \Lambda
_{-})+\overline{\Lambda }_{+}\Lambda _{+}h(hN-2g\Delta )\varphi \varphi
^{*}\right)  \nonumber \\
&&+\varphi \varphi ^{*}\left( 2h\Delta +2gh\overline{\Lambda }_{+}\Lambda
_{+}-g^2\partial _\mu N\partial ^\mu N\right)  \nonumber \\
&&\left. \left. -\frac h2(\varphi \overline{\Lambda }_{+}X_{-}+\varphi ^{*}%
\overline{X}_{-}\Lambda _{+})+\left| S+\frac g2\overline{X}_{-}\Lambda _{-}-%
\frac{g^2}2\varphi \overline{\Lambda }_{+}\Lambda _{-}\right| ^2\right]
\right\} ,  \label{canonic}
\end{eqnarray}
where now
\begin{equation}
\nabla _\mu \varphi =\left( \partial _\mu +ihA_\mu +ig\tilde{F}_\mu \right)
\varphi .
\end{equation}
Note that the last transformation changes only the kinetic and
topological-mass terms, but not the interaction terms.


\section{General conclusions}

We have here presented an N=1 Maxwell-BF model with nonminimal coupling
between matter and a 2-form gauge potential, as a supersymmetric version of
the CSKR model. As a by-product, we have obtained an N=2--susy extension of
a MCS system with nonminimal magnetic moment interactions after dimensional
reduction from D=4 to D=3.

Concerning the results of ref.\cite{navratil}, where the author extends
supersymmetry in the absence of a neutral scalar superfield, we remark that
our procedure provides an N=2--model containing such an ``extra'' scalar, $N$%
, as a natural consequence of the dimensional reduction. Thus, its presence
is well-justified: it appears as a three-dimensional descent of the D=4
gauge sector.

Also in contrast to the results of ref.\cite{navratil}, we do not need to
impose that all fields have the same mass. The reason is that, in our case,
we build up the N=2--D=3 action with independent matter and gauge
multiplets. The mass degeneracy then takes place only inside each multiplet.
In ref. \cite{navratil}, in turn, the N=2--D=3 model is formulated in terms
of a single gauge multiplet that encompasses the physical scalar among its
components. We claim that we place ourselves in the appropriate context for
generating {\it topological} vortices \cite{mha}, an issue which has not
still been thoroughly investigated.

Our construction is performed in terms of D=3--component fields. It would be
also interesting to carry out the dimensional reduction while working in
superspace, namely, to dimensionally reduce the N=1--D=4 superspace action
without passing through components. The D=3 superspace action must be
manifestly N=2 supersymmetric and its component-field projection is to be
compared with the action as written in eq. (\ref{canonic}). The three
dimensional version of the model may also be written in terms of
N=1--superfields. The results of our efforts in this direction will be soon
reported elsewhere \cite{inprog1}.

Finally, as a consequence of a suitably defined N=2 extension like (\ref
{canonic}) one can find out the proper Higgs potential allowing self-dual
vortex configurations \cite{sbobo}.

\section*{Acknowledgements}

The authors are grateful to O. Piguet, O.M. del Cima, D.H.T. Franco and M.A.
de Andrade for useful suggestions and discussions. Thanks are also due to
CNPq-Brazil, CAPES-Brazil, CLAF and FAPERJ-Rio for their fellowships.

\end{document}